\documentstyle[12pt]{article}

\parskip=\medskipamount			
\def\7#1#2{\mathop{\null#2}\limits^{#1}}        
\def\greaterthansquiggle{\raise.3ex\hbox{$>$\kern-.75em\lower1ex\hbox{$\sim$}}}
\def\lessthansquiggle{\raise.3ex\hbox{$<$\kern-.75em\lower1ex\hbox{$\sim$}}}
\def\Nsquiggle{\raise.3ex\hbox{$N$\kern-.75em\lower1ex\hbox{$\,\sim$}}}

\begin{document}
\def\thefootnote{\fnsymbol{footnote}}
\hfill{UMD-PP-94-160}
\begin{center}
{~~~}\\[.2in]
{\large\bf {Hints of Grand Unification in Neutrino Data
\footnote{ Work supported by a grant from the National Science Foundation}
} }\\[.5in]
{\large\bf { R.N.Mohapatra} }\\[.2in]
{\it Department of Physics,University of Maryland,College Park,Md.20742}\\
\end{center}
\begin{center}
{\bf Abstract}\\
\end{center}

     There are strong indications for neutrino masses and mixings
in the data on solar neutrinos ,  the observed deficit
of muon neutrinos from the atmosphere as well as from discussions
of the dark matter of the universe after COBE data.
It is argued that an SO(10) grand unified theory has
the right symmetry breaking properties needed to accomodate
 the neutrino masses and mixings suggested by these experiments. The minimal
version of the  model in fact
 leads to a {\it complete prediction} for the neutrino
masses and mixings  which can accomodate the observations partially, making the
theory testable in near future. If the model is supplemented by
an $S_4$ horizontal symmetry, it leads to a highly degenerate light
neutrino spectrum which is the only way fit all data with the three
known light neutrinos.

\begin{center}
{\bf I.\underline{Introduction}:}
\end{center}

   One of the strogest indications of new physics beyond
the standard model is
in the arena of neutrinos where there are experimental
results, which can be understood most easily if the
neutrinos are assumed to have  nonvanishing
 masses and mixings . The experimental results are:
  i) the deficit of solar neutrinos now observed in four
different experiments[1] compared to the calculations
based on the standard solar model[2]
 ii) the depletion of atmospheric muon
neutrinos observed in three different experiments[3]
compared to calculations[4]; and iii)
the apparent need for some hot dark matter in the Universe[5].
In this talk, I will first argue that the masses and mixings
for neutrinos required by the above data are very strongly
suggestive of an SO(10) grandunified theory beyond the standard
model; then I
present a recent work with
K. S. Babu[6] which showed that the solar neutrino
results are most easily accounted for by the minimal SO(10)
grandunified theory where the constraints of grandunification
and a realistic charged fermion spectrum allow a complete
prediction for both the neutrino masses and mixings upto an
overall scale . This minimal
model however cannot accomodate the atmospheric neutrino data in
conjunction with the solar neutrino data.
Caldwell and I [7] have argued that the only way to accomodate
all the above neutrino observations
 with only three light neutrinos is to have a degenerate
spectrum for them, a property that can emerge in an
  $SO(10)$ model, if it is supplemented by extra horizontal
symmetries[7,8,9].

 Let us first summarize the values
for neutrino masses and mixings required to understand the data
on the basis of simple two neutrino mixing.

{\it 1.1 Solar Neutrino Deficit:}

 In the two neutrino mixing approximation,
 the following choice of masses and mixings is consistent with present
data[10]:
i) The small angle non-adiabiatic MSW solution[11]:
$\Delta m^2_{\nu_e \nu_i}\simeq (.3-1.2)\times 10^{-5} eV^2$
 and $ sin^2{2\theta}\simeq (.4-1.5)\times 10^{-2}$
ii) Large angle MSW solution:
$\Delta m^2_{\nu_e\nu_i}\simeq (.3-3)\times 10^{-5}eV^2$
 and $sin^2{2\theta}\simeq .6-.9$
iii) Vacuum oscillation solution:
$\Delta m^2_{\nu_e\nu_i}\simeq(.5-1.1)\times 10^{-10}eV^2$
 and $sin^2{2\theta}\simeq(.8-1)$

{\it 1.2 Atmospheric Neutrino Puzzle:}

 A straightforward way to understand the deficit of the muon
neutrinos is to assume
 that $\nu_\mu$ oscillates to another light neutrino.Assuming
 the latter to be the tau neutrino,
the data can be fitted with[4] the values of $\Delta m^2_{\nu_\mu
\nu_{\tau}}\simeq .5-.005 eV^2$ and $sin^2{2\theta}\simeq .5$.
We do not consider the alternative possibility that atmospheric
neutrino anomaly
could be resolved via $\nu_{\mu}$-$\nu_e$ oscillation. Although strictly
this is not ruled out[4,12], it would imply distortion in the observed
$\nu_e$ spectrum in the underground experiments for which there seems
to be no evidence.

{\it 1.3 Hot Dark Matter Neutrinos:}

   Data on the extent of structure in the universe
available on a wide range of distance scales together with
the COBE results on the anisotropy of the cosmic microwave
background radiation, galaxy-galaxy angular correlation ,
 large scale velocity fields, and correlations of galactic
clusters can all be fit[5] by a model of the universe
containing $70\%$ cold dark matter and $30\%$ hot dark matter.
( But perhaps an admixture in the ratio  $90\%$ to $10\%$
of CDM to HDM may not be inconsistent ). Although, there are
other possibilities such as using the cosmological constant
in conjunction with CDM, tilted spectrum plus CDM etc.,the
mixed dark matter scenerio has its own appeal since the already
known neutrino with mass in the appropriate range of $7 eV~ to~ 2 eV$
could be the HDM.

{\it 1.4 Possible signal of an eV Majorana mass in $\beta\beta_{0\nu}$
decay }

More recently the data from $^{76}Ge$[13] and $^{130}Te$[13]
 neutrinoless double
beta decay($\beta\beta_{0\nu}$) experiments have led to the
possibility that the existence of
 an effective Majorana mass,$<m_\nu>\simeq{1-2}eV$ can either
be confirmed or ruled out in very near future.

 There is of course a tentativeness to some of the data under
consideration. Nevertheless, we believe it is not premature to
discuss what their implications are for neutrino mass matrices
and physics beyond the standard model.

\bigskip

\begin{center}
{\bf II.\underline{Neutrino Mass Matrices Suggested by Data}:}
\end{center}
\bigskip

   The kind of neutrino spectrum
and their mass matrices that would be required to fit the above
observations has been the subject of two recent papers by D.Caldwell
and this author[7]. We have found two possible scenerios , which
fit all the above constraints; but the most economical one that
uses only the three known light neutrinos has a very intriguing
structure that we give below. The  $\nu_e$, $\nu_\mu$ and
$\nu_\tau$ are all nearly degenerate with mass around 2
eV. The mass differences are appropriately arranged so that
$\nu_\mu$-$\nu_\tau$ oscillations explain the atmospheric neutrino
problem and similarly $\nu_e$ - $\nu_\mu$ mass differences as well
as mixings are so arranged that they can explain the solar neutrino
deficit via the  MSW mechanism mechanism using the small angle
non-adiabatic solution. The simplest mass matrix, which can achieve
this is:

\renewcommand\arraystretch{1.1}
$$ M = \left( \begin{array}{ccc}
 m+\delta_1 s^2_1 & -\delta_1 s_2 c_1 c_2  & -\delta_1 c_1 s_1 s_2 \\
-\delta_1 s_2 c_1 c_2  &  m+\delta_1 c^2_1 c^2_2 + \delta_2 s^2_2  &
(\delta_1  -\delta_2) s_2 c_2 \\
-\delta_1 c_1 s_1 s_2  & (\delta_1 -\delta_2) c_2 s_2  &
 m+\delta_1 s^2_2 +\delta_2 c^2_2
\end{array} \right)$$     \hfill (1)

In eq.(1), $m\simeq 2 eV$; $\delta_1\simeq 1.5\times 10^{-6} eV$;
$\delta_2\simeq .2~ to~ .002 eV$; $s_1\simeq .05$
and $s_2\simeq .35$. It is worth repeating that
Majorana mass for $\nu_e$ of this magnitude will be tested by the
current generation of neutrinoless double beta decay experiments.
 Obviously, hot dark matter in thus case is, distributed
between the three active species of neutrinos almost equally.

 In view of the tentative nature of some of the data at the moment
, first we explore the theoretical
implications of  a non-vanishing
neutrino mass in the simplest grand unified model based on the group
SO(10) so that the simplest model can be exposed
to tests via neutrino experiments . Then I discuss what modifications
are needed to fit all data in the SO(10) framework.

\bigskip
\begin{center}
{\bf III.\underline{ Massive Neutrinos , Local B-L Symmetry and
SO(10) Grandunification }:}
 \end{center}
\bigskip

    Let us start by reminding the reader that in the standard model,
the neutrinos are massless because only the lefthanded
 neutrinos appear in the the  spectrum
and  $B-L$ is an exact symmetry of the Lagrangian. In order
to obtain massive neutrinos, one must therefore include the right-handed
neutrino in the spectrum. It however turns out that as soon as this
is done, in the theory there appears a completely triangle anomaly free
generator, the $B-L$. This symmetry is then a gaugeable symmetry and
it would be rather peculiar if nature chooses not to gauge a symmetry
which is gaugeable. If following this line of reasoning, we use $B-L$
as a gauge symmetry, the most natural gauge group turns out to be
the Left-Right symmetric group $SU(2)_L\times SU(2)_R\times U(1)_{B-L}$ [14],
which breaks at some high scale to the standard model group.
The quarks and leptons in this model are assigned in a completely left-right
symmetric manner, i.e. if we define $Q\equiv (u,d)$ and $\psi \equiv (\nu,
e)$, then $Q_L(2,1,1/3)$ and $Q_R(1,2,1/3)$ are assigned in a left-right
symmetric manner and similarly , $\psi_L(2,1,-1)$ and $\psi_R(1,2,-1)$.
The Higgs sector of the model that leads naturally to small
neutrino masses in this model consists of the bi-doublet field
$\phi\equiv (2,2,0)$ and the triplet fields $\Delta_L\equiv (3,1,+2)$
and  $\Delta_R\equiv (1,3,+2)$ [15]. The Yukawa couplings of
the model are :
$$L_Y = h_1 \bar{Q}_L\phi Q_R + h_1^{\prime}\bar{Q}_L \tilde{\phi} Q_R
h_{\ell}\bar{\psi}_L \phi \psi_R + h^{\prime}_{\ell}\bar{\psi}_L
\tilde{\phi} \psi_R + f\psi^T_L C^{-1}\tau_2 \Delta_L\psi_L + L\rightarrow
R  + h.c. \;\eqno(2)$$

  The gauge symmetry breaking is achieved in two stages : in the first
stage, the neutral component of $\Delta_R$ multiplet acquires a vev
$<\Delta^0_R>=v_R$, thereby breaking the gauge symmetry down to
the $SU(2)_L\times U(1)_Y$ group of the standard model; in the
second stage, the neutral components of the multiplet $\phi$
acquire vev breaking the standard model symmetry down to $U(1)_{em}$.
At the first stage of symmetry breaking, $W_R$ and $Z^{\prime}$
acquire masses of order $gv_R$ and in the second stage the familiar
$W_L$ and $Z_L$ acquire masses. The near maximality
 of parity violation at low energies is due to
the masses of $W_R$ and $Z^{\prime}$ being bigger than those of the
$W_L$ and the $Z$ boson. At the first stage of symmetry
breaking, the f-terms in the Yukawa coupling give nonvanishing masses
to the three right-handed neutrinos of order $fv_R$ keeping all other
fermions massless. At the second stage , quarks, charged
leptons as well as
the neutrinos acquire Dirac masses. The $\nu_L$-$\nu_R$ mass mass
matrix at this stage is a $6\times 6$ mass matrix of the following
see-saw form[16]:
\renewcommand\arraystretch{1.1}
$$   M_{\nu}=\left(\begin{array}{cc}
             0 &  m_D \\
             m^T_D& M_R
      \end{array}\right)    \;\eqno(3)$$

As is well known this see-saw form leads to three light eigen-values
generically of order
$$m_{\nu_{i}}\simeq -\left( {m_D} {M_R}^{-1} {m^T_D} \right)\;\eqno(4)$$

The typical values of $m_D$ are expected to be of order of the charged
fermion masses in the theory whereas the $M_R$ corresponds to the scale
of $B-L$ breaking which is a very high scale, thereby explaining the
smallness of the neutrino masses. The specific value of $m_D$ is however
model-dependent and depending on what the value of $m_D$ is, the
spectrum of the light left-handed Majorana neutrinos will be of the
eV : keV : MeV type or of the micro : milli : eV type. The former type
of spectrum can be tested in the double beta decay as well as the
conventional beta decay end-point experiments whereas the second
spectrum can be tested in the solar neutrino as well as the long
base-line neutrino experiments.

In view of the discussion of the previous section, the micro-milli-eV
spectrum for the light neutrinos
is of great current interest. In the simple see-saw models that
naturally emerge in the left-right symmetric models, one generically
has  $m_D\simeq m_f$,where $f=$ leptons or quarks; so
if we  want $m_{\nu_{\mu}}\simeq 10^{-3}$eV,then the mass
$M_R$ must be of order $10^{10}-10^{12}$GeV. This would suggest
grandunification models of type $SO(10)$ or some higher group
containing it. The $SO(10)$ possibility is the most exciting
because all its symmetry breaking scales i.e. the GUT scale
$M_U$ and the $B-L$ breaking scale $M_R$ are predicted by the LEP data
and amazingly enough, they are precisely in the above mantioned
range for the non-SUSY versions of the model[17]. Such an intermediate
scale is also required for adequate cosmological baryogenesis[18] and
the tau neutrino being the hot dark matter of the Universe.
This constitutes enough circumstantial evidence to take the SO(10)
model seriously and study its detailed predictions so that it can
be subjected to experimental testing.

\newpage
\begin{center}
{\bf IV.\underline{ Minimal SO(10) GUT and Predictions
 for Neutrino Masses and Mixings}:}
\end{center}
\bigskip

   As we saw in the previous section, the simple see-saw
model predicts a scale of B-L symmetry breaking near $10^{11}$
GeV or so if it is to solve the solar neutrino puzzle. Both
the see-saw formula as well as  a large $B-L$ symmetry scale
emerge naturally from the SO(10) models. The minimal $SO(10)$
model without supersymmetry leads to a two step breaking of $SO(10)$
down to the standard model. There are four possibilities, two
corresponding to the case where the discrete $Z_2$ local subgroup
( called D-parity)[19] is broken and two where D-parity survives
down to the $B-L$ breaking scale.
In the D-parity broken case, we  have the intermediate
symmetry group to be
 $SU(2)_L\times SU(2)_R\times G_c$
where $G_c$ is $SU(4)_C$ (denoted as case (A))or
 $SU(3)_c\times U(1)_{B-L}$ ( denoted as case (B) ).
The advantage of this case is that it makes the conventional
 see-saw formula natural[20].
Use of Higgs
multiplets belonging to {\bf 210} and {\bf {45}+{54}}
representations to break SO(10)
leads to such a scenerios (A) and (B) respectively. It however turns out that
in order to realize the degenerate neutrino spectrum, one needs to preserve
D-parity down to the scale of $B-L$ symmetry breaking where one has to use
the second
two possibilities. Depending on whether the color gauge subgroup
below GUT scale is  $SU(4)_c$ or $SU(3)_c\times U(1)_{B-L}$; we denote these
cases as case (C) and case (D) respectively.

  A very important point worth emphasizing here is that inputting the
LEP data for the three gauge couplings for the standard model leads to
unique predictions for the unification scale $M_U$ and the
intermediate scale $M_I$.
These predictions for
non-SUSY version of the model
have been studied including two-loop and threshold corrections in ref.17
and 21 and the results are:
$$Model (A):~~~~~M_U = 10^{16.26^{+.13}_{-1.24}\pm .25} GeV
{}~~~~~~~~~M_I = 10^{10.7^{+2.65}_{-.07}\pm .02} GeV \;\eqno(5)$$
$$Model (B):~~~~~M_U = 10^{16.42\pm .18 \pm .25} GeV
{}~~~~~~~~~M_I = 10^{9^{+.69}_{-.3} } GeV \;\eqno(6)$$
$$Model (C):~~~~~M_U~=~10^{15.02\pm .48 \pm .25}~GeV
{}~~~~~~~~~M_I~=~10^{13.64 \pm .88}~GeV \;\eqno(7)$$
$$Model (D):~~~~~M_U~=~10^{15.55\pm .43\pm .20}~GeV
{}~~~~~~~~~M_I~=~10^{10.16\pm .57}~GeV \;\eqno(8)$$

 First ,we note that the values of the intermediate scale are in
the range required by the see-saw formula to give the neutrino masses
which can play a role in the understanding of the various anomalies
described in the introduction. Whether they really do or not
depends of course on the various mixing angles. We will see that
in the minimal models, the mixing angles are in the right range
(contrary to a common belief in some quarters that the neutrino mixing
angles should mirror the quark CKM mixing angles ) .

Secondly, we also have prediction for the proton life-time in non-SUSY
$SO(10)$ models for these cases:

$$\tau_p = 1.44\times~10^{37.4\pm .7\pm 1.0 ^{+.5}_{-5}} years~~~ Model (A)$$
$$\tau_p = 1.44\times~10^{37.7 \pm .7\pm .9^{+.5}_{-2.0}} years~~~~Model(B)$$
$$\tau_p = 1.44\times~10^{32.1\pm .7\pm 1.0\pm 1.9} years~~~~~Model(C)$$
$$\tau_p = 1.44\times~10^{34.2\pm .7\pm .8\pm 1.7} years~~~~~Model (D)$$

Some of these predictions are within
 the reach of the Super-Kamiokande experiment[22],
which should therefore throw light on the non-SUSY version of the
$SO(10)$ model.

Let us now discuss the predictions for neutrino masses in the
minimal SO(10) model. This
necessitates  detailed knowledge
 of the Dirac neutrino mass matrix as well as the Majorana neutrino mass
matrix.  Luckily, it turns out that in $SO(10)$ models,
the charge $-1/3$ quark mass matrix is
related to the charged
lepton matrix and
the neutrino Dirac
mass matrix is related
to the charge $2/3$ quark matrix at the unification scale.
However, prior to the work of ref.6 , no simple way was known
 to relate the heavy Majorana matrix to the
charged fermion observables.  This stood in the way of predicting the
light neutrino spectrum. It was however shown in ref.[6] that
 in a class of minimal $SO(10)$
models, in fact, not only the Dirac neutrino matrix, but the Majorana
matrix also gets related to observables in the charged fermion sector.
This leads to a very predictive neutrino spectrum .
We use a simple Higgs system with
one (complex) {\bf 10}
and one {\bf 126} that have Yukawa couplings to fermions.
The {\bf 10} is needed for
quark and lepton masses, the
{\bf 126} is needed for the see--saw mechanism.  Crucial to the
predictivity of the neutrino spectrum is the observation that
the standard model doublet contained in the {\bf 126}
receives an induced vacuum expectation
value (vev) at tree--level.  In its absence, one would
have the asymptotic mass
relations $m_b=m_\tau,~m_s=m_\mu,~m_d=m_e$.
While the first relation would lead to
a successful prediction of $m_b$ at low energies, the last two
are in disagreement with observations.  The induced vev of the standard
doublet of {\bf 126} corrects these bad relations and at the
same time
also relates the Majorana neutrino mass matrix to
observables in the charged fermion sector, leading to a predictive
neutrino spectrum.

We shall consider non--Susy $SO(10)$ breaking to the standard model via
the $SU(2)_L \times SU(2)_R
\times SU(4)_C \equiv G_{224}$ chain as well as
Susy-$SO(10)$ breaking directly to the standard model.
The breaking of $SO(10)$ via $G_{224}$ is achieved by
a {\bf 210} of Higgs which breaks
the discrete $D$--parity.
The second stage of symmetry breaking
goes via the {\bf 126}.  Finally, the electro--weak symmetry breaking
proceeds via the {\bf 10}.
In Susy-$SO(10)$, the
first two symmetry breaking scales coalesce into one.

In the fermion sector, denoting
the three families belonging to
{\bf 16}--dimensional spinor representation of $SO(10)$ by
$\psi_a$, $a=1-3$, the complex {\bf 10}--plet of Higgs by $H$, and
the {\bf 126}--plet of Higgs by $\Delta$,
the Yukawa couplings can be written down as
$$
L_Y = h_{ab}\psi_a\psi_bH + f_{ab}\psi_a\psi_b\overline{\Delta} + H.C.
\;\eqno(9)$$
Note that since the {\bf 10}--plet is complex, one other coupling
$\psi_a\psi_b\overline{H}$ is allowed in general.  In Susy--$SO(10)$, the
requirement of supersymmetry prevents such a term.  In the non--Susy
case, we forbid this term by imposing a $U(1)_{PQ}$
symmetry, which may anyway be needed in order to solve the strong CP
problem.

The {\bf 10} and {\bf 126} of Higgs have the following decomposition
under $G_{224}$:
${\bf 126} \rightarrow (1,1,6)+(1,3,10)+(3,1,\overline{10})+(2,2,15)$,
${\bf 10} \rightarrow (1,1,6)+(2,2,1)$.
Denote the $(1,3,10)$ and $(2,2,15)$ components of
$\Delta({\bf 126})$
by $\Delta_R$ and
$\Sigma$ respectively and the $(2,2,1)$ component of $H({\bf 10})$ by
$\Phi$.
The vev $<\Delta_R^0> \equiv v_R \sim 10^{12}~GeV$
breaks the intermediate symmetry down to
the standard model and generates Majorana
neutrino masses given by $fv_R$.
$\Phi$  contains two standard model doublets
which acquire
vev's denoted by $\kappa_u$ and $\kappa_d$ with
$\kappa_{u,d} \sim 10^{2}~GeV$.
$\kappa_u$ generates charge 2/3 quark as well as Dirac neutrino
masses, while $\kappa_d$ gives rise to $-1/3$ quark and charged lepton
masses.

Within this minimal picture, if $\kappa_u,~\kappa_d$ and $v_R$ are
the only vev's
contributing to fermion masses, in addition to
the $SU(5)$ relations $m_b=m_\tau,~
m_s=m_\mu,~m_d=m_e$, it will also lead to the unacceptable relations
$m_u:m_c:m_t = m_d:m_s:m_b$.
Moreover, the
Cabibbo-Kobayashi-Maskawa (CKM) mixing matrix will be identity.
It was however shown in ref.6, that in this model there exist
new contributions
to the fermion mass matrices which are of the right order of
magnitude to correct these bad relations.  To see this, note that
the scalar potential contains, among other terms, a crucial term
$V_1 =\lambda \Delta \overline{\Delta} \Delta H +H.C.$
Such a term is invariant under the $U(1)_{PQ}$ symmetry.  It will be
present in the Susy $SO(10)$ as well, arising from the {\bf 210}
$F$--term.
This term induces vev's for the standard doublets contained in the $\Sigma$
multiplet of {\bf 126}.  The vev arises through a term
$\overline{\Delta}_R\Delta_R \Sigma \Phi$ contained in $V_1$.
 The magnitudes of the induced vev's
of $\Sigma$ (denoted by $v_u$ and $v_d$ along the up
and down directions) can be estimated using  the survival hypothesis :
$
v_{u,d} \sim \lambda \left({v_R^2}/ {M_{\Sigma}^2}\right)
\kappa_{u,d}~~.$
Suppose $M_U \sim 10^{15}~GeV$, $M_I \sim 3 \times 10^{12}~GeV$
and $M_{\Sigma} \sim 10^{14}~
GeV$, consistent with survival hypothesis, then $v_u$ and
$v_d$ are of order 100 MeV, in the right range for correcting the bad
mass relations.  We emphasize that there is no need for a
second fine--tuning to generate such induced vev's.  In the Susy
version with no intermediate scale, the
factor $(v_R^2/M_{\Sigma}^2)$ is not a suppression, so
the induced vev's
can be as large as $\kappa_{u,d}$.

We are now in a position to write down the quark and lepton mass
matrices of the model:
$$
M_u = h \kappa_u +fv_u~~$$
$$M_d = h \kappa_d+f v_d$$
$$M_{\nu}^D = h \kappa_u-3 f v_u$$
$$M_l=h \kappa_d-3fv_d$$
$$M_{\nu}^M = f v_R~.
\;\eqno(10)$$
Here $M_{\nu}^D$ is the Dirac neutrino matrix and $M_{\nu}^M$ is the
Majorana mass matrix. Let us ignore CP-violation , which has been
taken into account in [6]. Note that, there are
12 parameters in all, not counting the superheavy scale $v_R$: 3
diagonal elements of the matrix $h \kappa_u$, 6 elements of $f v_u$,
and three vev's. These are completely determined by the
charged fermion sector, viz., 9 fermion masses, 3 quark mixing
angles .  The light neutrino mass matrix is then completely predicted
upto the overall scale $v_R$. In making the predictions, we have been
careful to take into account the renormalization extrapolation of
the relations in eq. to the weak scale.
Below, we present results for the non--Susy $SO(10)$ model with the
$G_{224}$ intermediate symmetry.  We fix
the intermediate scale at $M_I = 10^{12}~GeV$.
We find that there are essentially three different solutions. The one
that can fit the solar neutrino data is the one below.
$$
{\rm Input}: m_u(1~GeV) = 3~MeV,~~m_c(m_c)=1.22~GeV,~~m_t=150~GeV $$
$$m_b(m_b) = -4.35~GeV,~~r_1=-1/51,~~r_2=0.2 $$
$${\rm Output}: m_d(1~GeV) = 5.6~MeV,~~ m_s(1~GeV)=156~MeV $$
$$\left(m_{\nu_e},m_{\nu_\mu},m_{\nu_\tau}\right) = R\left(7.5 \times
10^{-3},2.0,-2.8 \times 10^3\right)~GeV $$
\renewcommand\arraystretch{1.1}
$$V_{KM}^{\rm lepton} = \left(\begin{array}{ccc}
0.9961 & 0.0572 & -0.0676 \\
-0.0665 & 0.9873 & -0.1446 \\
 0.0584 & 0.1485 & 0.9872 \end{array} \right)$$.
\hfill (11)

where $R=v_u/v_R$.

Note that the pattern of mixing angles is very different from the
quark sector and both the $\nu_e$-$\nu_{\mu}$ and $\nu_e$-$\nu_{\tau}$
mixing angles are in the range to be useful in understanding the
solar neutrino puzzle. Moreover, the
$\nu_\mu-\nu_\tau$ mixing angle is  near $3 |V_{cb}|$, so that
the present neutrino oscillatin data[23] implies that
  $m_{\nu_{\tau}} \le 2~ eV$.  From the $\nu_\tau/\nu_\mu$ mass
ratio, which is $1.4 \times 10^3$ in this case, we see that
$m_{\nu_\mu} \le 1.5 \times 10^{-3}~eV$.  This is just within the
allowed range[10]  for small angle non--adiabatic
$\nu_e-\nu_\mu$ MSW oscillation, with a predicted count rate of about
50 SNU for the Gallium experiment.
Note that there is a lower limit of about 1 eV for
the $\nu_\tau$ mass in this case.  Forthcoming experiments
( CHORUS and NOMAD[24] at CERN and the Fermilab expt.) should then
be able to observe $\nu_\mu-\nu_\tau$ oscillations.  A $\nu_\tau$ mass
in the (1 to 2) eV range can also be cosmologically significant, it can
be at least part of the hot dark matter.

 Three more sets of predictions for neutrino masses and mixings
in this model have been found by Lavoura[25]; none of them have
features needed to accomodate both the solar and the atmospheric neutrino
puzzle.

Before closing this section, let me make some comments on the SUSY-SO(10)
model. First, if the minimal model discussed is supersymmetrized, the
predictions for neutrino masses and mixings remain unchanged
- with the
difference that the $B-L$ scale which appears in the overall coefficient
in the neutrino mass matrix
is now same as the GUT scale. So, the  more natural possibility here
is to solve the solar neutrino puzzle via the $\nu_e$-$\nu_{\tau}$
oscillation since due to the high value of $v_R$, it is the
tau neutrino mass which is more easily of order $10^{-3}$ eV.
 It is however possible that with the
 inclusion of threshold corrections, the $B-L$ symmetry
breaking scale is somewhat lower than the GUT scale and
the muon neutrino remains as milli-eV particle
still allowing the $\nu_e$-$\nu_{\mu}$ oscillation
solution to the solar neutrino puzzle.
It is also important to point out that the SUSY SO(10)
has the advantage that it automatically provides a cold dark
matter candidate, the lightest supersymmetric particle (the LSP)
due to the fact that R-parity is an automatic symmetry of the
model. In the non-susy models we have to invoke perhaps an
axion as the CDM[26]. In both models, there appears to be no
HDM candidate unless a two eV tau neutrino is considered
adequate by cosmologists for the purpose.

\bigskip
\begin{center}
{\bf V.\underline{An SO(10) model for a degenerate neutrino scenario:}}
\end{center}

 In this section, we discuss the ingredients needed to build a model
for degenerate neutrinos of the type discussed in section II in order
to fit all the data summarized in sec.I.
 The basic strategy is to employ the
  fact that when the conventional see-saw mechanism for
neutrino-masses is implemented
 in gauge models such as SO(10) or the
left-right symmetric models,
 it gets modified to the following form[15]
$$
\left( \begin{array}{cc}
 f v_{L} & m_{v^{D}} \\  m_{v^{D}}^{T} & f v_{R}

 \end{array} \right), \;\eqno(12)$$
where
$v_{L}=\lambda {v_{wk}^2 v_{R} \over M_{P}^2 }$;
$v_{R}$ is the scale of SU(2)$_{R}$-breaking
 and $M_{P}$ is breaking scale of parity.
Therefore, unless special care is taken
 to break parity symmetry at a scale higher than the
SU(2)$_{R}$ or U(1)$_{B-L}$, $v_{L}\sim \lambda v_{wk}^2 / v_{R}$ (since
$v_{R}\sim
M_{P}$ ).
The light neutrino masses are then given by:
$$m_{\nu}\simeq fv_{L}-{m_{\nu^D} f^{-1} m_{\nu^D}^T \over v_{R} }.
\;\eqno(13)$$

Recall that the conventional see-saw formula omits the first term (which
 is justified only under special circumstances) .
 We will however keep both the terms in the present discussion.
Now notice that if due to some symmetry reasons,
 $f_{ab}=f_{0} \delta_{ab}$, then a degenerate
neutrino spectrum emerges. This property has been used in several
recent papers[7,8,9,27] to obtain a nearly degenerate spectrum
for light neutrinos. In the rest of the paper, we discuss the
model given in ref.8.

 Consider the breaking of SO(10) $\rightarrow$ SU(2)$_L$
$\times$ SU(2)$_R$ $\times$ SU(4)$_C$ $\times$ P (denoted by $G_{224P}$) by
means of
a $\{{\bf 54}\}$-dim. Higgs multiplet. This symmetry is subsequently broken
down to the
standard model by a $\{{\bf 126}\}$-dim. Higgs multiplet. Detailed two-loop
analysis of the
mass scales in this model[21] leads to $v_{R}\sim 10^{13.6}$ GeV. So that for
$f_{0}
\lambda \sim 1/2$, we get
$f_{0}v_{L}\sim 1$ eV, as desired. We will supplement this model by a
softly broken $S_4$ symmetry which restricts the Yukawa couplings in
such a way that it not only leads to realistic charged fermion
masses but also to the following predictions for the neutrino
masses and mixings[8].

Writing $m_{\nu_i}=m_0+
m_{\nu_i}^{'}$, where $m_0\simeq 2 eV$
 is the direct $v_L$ contribution, we give
a set of predictions for the masses and mixing angles which fit all
known observations:
$(m_{\nu_{e}}^{'}, m_{\nu_{\mu}}^{'}, m_{\nu_{\tau}}^{'})=
{1 \over  f v_R} (-0.0000174465,
-0.129248, -5759.27 ) GeV^2$
$$V^l = \left( \begin{array}{ccc}
-.9982 & .05733 & .01476 \\ .05884 & .9334 & .3541 \\ -.006523 & -.3544 & .9351
\end{array} \right)\;\eqno(14)$$

Note that, for $v_R \simeq 10^{13.6}$ GeV and $f \sim 3$, this predicts
$|m_{\nu_{\mu}}^{2}-m_{\nu_e}^{2}| \sim 4 \times 10^{-6}$ eV$^2$ for $m_0=$ 2
eV,
$|m_{\nu_{\tau}}^{2}-m_{\nu_{\mu}}^{2}| \sim .2$ eV$^2$, which are
 in the range required to
solve both the solar and atmospheric neutrino deficit for the values of
 $\theta_{\nu_{e}
\nu_{\mu}}$ and $\theta_{\nu_{\mu} \nu_{\tau}}$ given above.
In particular, we wish to note the preference of theory for the
small angle MSW solution to the solar neutrino problem.

\bigskip
\begin{center}
{\bf VI.\underline{ Summary and Conclusions}:}
\end{center}
\bigskip

In summary, I have argued in this report that if the present data
on solar neutrinos and the C+HDM picture of the universe
are taken seriously, then
the most natural theoretical framework to understand their implications
for neutrino masses and mixings is an
SO(10) GUT models with the  right-handed
scale ( or $B-L$ breaking scale )
  in the super-heavy range of $10^{11}$ GeV or so. This result becomes
more compelling, once one realizes that precisely such a value for
the $B-L$ scale is implied by the low energy LEP data  applied to
a non-SUSY SO(10) model. Furthermore, in
 the minimal version of the  SO(10) model, the values for
neutrino masses and mixings are completely predicted and they
 fit the solar neutrino data rather beautifully and predict a tau
neutrino mass around 2 eV. This is a bit low to be a good hot dark
matter candidate but its role as a weak HDM may not be ruled out.
The predictions for the mixing angle in the $\nu_e$-$\nu_{\tau}$
sector
can be tested by the neutrino oscillation experiments such as CHORUS, NOMAD
and the Fermilab experiments
and proton decay searches to be carried out at SuperKamiokande.
 In fact, the present
atmospheric neutrino data cannot be accomodated by the minimal $SO(10)$
model; therefore if this data stands the test of time, a second minimal
grandunified model will be ruled out by experiments and one may be forced
into a degenerate neutrino scenario described in sec.V above.
I then discuss, how the degenerate scenario may emerge in an SO(10)
GUT framework.

\newcounter{000}
\centerline{\bf References}
\begin{list}{[~\arabic{000}~]}
{\usecounter{000}\labelwidth=1cm\labelsep=.5cm}

\item R. Davis et. al., in {\it Proceedings of the 21st International
Cosmic Ray Conference}, Vol. 12, ed. R.J. Protheroe (University of
Adelide Press, Adelide, 1990) p. 143; \\
K.S. Hirata et. al., {\it Phys. Rev. Lett.} {\bf 65}, 1297
(1990); \\
A.I. Abazov et. al., {\it Phys. Rev. Lett.} {\bf 67}, 3332 (1991); \\
P. Anselman et. al., {\it Phys. Lett.}
{\bf B285}, 376 (1992);\\
T. Bowles,Invited talk at ICNAPP at Bangalore, January (1994);\\
T. Kirsten, Invited talk at ICNAPP at Bangalore, January (1994).
\item J. Bahcall and M. Pinnsonnault, {\it Rev. Mod. Phys.}
{\bf 64}, 885 (1992);\\
S. Turck-Chieze and I. Lopez, {\it Ap. J. } {\bf 408}, 347 (1993).
\item KAMII.K. Hirata et al.,{\it Phys. Lett.}{\bf B280}, 146 (1992);\\
IMB: R. Becker-Szendy et al., {\it Phys. Rev.}{\bf D 46}' 3720 (1992);\\
Soudan II: M. Goodman , APS-DPF Meeting, Fermilab , November, 1992;\\
Frejus: Ch. Berger et al.,{\it Phys.Lett} {\bf B245}' 305 (1990);\\
Baksan: M.M.Boliev et al., in {\it Proc. 3rd Int. Workshop on
Neutrino Telescopes} (ed. M. Baldo-ceolin, 1991),p.235.
\item W. Frati, T. Gaisser, A. Mann and T. Stanev, {\it Phys. Rev.}
{\bf D48}, 1140 (1993);\\
For a recent review, see S. Pakvasa, Invited talk at the ICNAPP, January,
(1994).
\item E.L.Wright et al.,{\it Astrophys. J.}{\bf 396}, L13 (1992);\\
R. Schaefer and Q. Shafi, BA-92-28 (1992);\\
J. A. Holtzman and J. Primack, {\it Astrophys. J.}{\bf 405}, 428 (1993).
\item K. S. Babu and R. N. Mohapatra, {\it Phys. Rev. Lett.} {\bf 70},
2845 (1993).
\item D. Caldwell and R. N. Mohapatra, {\it Phys. Rev.} {\bf D48},3259 (1993)
 and {\it Phys. Rev.} {\bf D50} (1994) ( to appear).
\item D. G. Lee and R. N. Mohapatra, University of Maryland Preprint No
UMD-PP-94-95 (1994);{\it Phys. Lett.B} to appear.
\item A. Ioannyssian and J. W. Valle, Valencia Preprint (1994).
\item  See for instance, N. Hata and P. Langacker, Pennsylvania
Preprint UPR-0592T (1993);\\
P. Krastev and S. T. Petkov, {\it SISSA Preprint No.177/93/EP};\\
G. Fogli, E. Lisi and D. Montanino, CERN-TH.6944/93.
\item L. Wolfenstein, {\it Phys. Rev.} {\bf D 17}, 2369 (1978); \\
S.P. Mikheyev and A. Yu Smirnov, {\it Yad. Fiz.} {\bf 42}, 1441 (1985)
[{\it Sov. J. Nucl. Phys}, {\bf 42},
913 (1985)].
\item E. Akhmedov, A. Lipari and M. Lusignoli, {\it Phys. Lett.}
{\bf B300}, 128 (1993 );\\
A. Acker, S. Pakvasa and J. Pantaleone, {\it Phys. Lett.} {\bf B298},
149 (1993).
\item H. Klapdor-Kleingrothaus, {\it Prog. in Part. and Nucl. Phys.}
(1994) ( to appear);\\
 E. Garcia, Talk at TAUP'93.
\item J.C. Pati and A. Salam,{\it Phys.Rev.D$\,$}{\bf 10}, 275 (1974)\\
R. N. Mohapatra and J.C. Pati,{\it Phys.Rev.D$\,$}{\bf 11}, 566, 2558 (1975)\\
G. Senjanovic and R.N. Mohapatra,{\it Phys.Rev.D$\,$}{\bf 12}, 1502 (1975).
\item R. N. Mohapatra and G. Senjanovic , {\it Phys. Rev. Lett.}
{\bf 44} , 912 (1980) ; {\it Phys. Rev. }{\bf D23} , 165 (1981).
\item M. Gell-Mann, P. Ramond and R. Slansky, in {\it Supergravity}, ed.
F. van Nieuwenhuizen and D. Freedman (North Holland, 1979), p. 315; \\
T. Yanagida, in {\it Proceedings of the Workshop on Unified Theory and
Baryon Number in the Universe}, ed. A. Sawada and H. Sugawara, (KEK,
Tsukuba, Japan, 1979);\\
R.N. Mohapatra and G. Senjanovic {\it Phys. Rev. Lett.} {\bf 44}, 912
(1980).
\item D.Chang,R.N.Mohapatra,J.Gipson,R.E.Marshak and M.K.Parida,
 {\it Phys. Rev. D$\,$}{\bf 31}, 1718 (1985);\\
R. N. Mohapatra and M. K. Parida, {\it Phys. Rev. } {\bf D47}, 264 (1993);\\
N. G. Deshpande, R. Keith and P. B. Pal,{\it Phys. Rev.}{\bf D46}
,2261 (1992).
\item M. Fukugita and T. Yanagida, {\it Physics of  Neutrinos}, to
be published by Springer-Verlag (1994).
\item D. Chang, R.N. Mohapatra and M.K. Parida, {\it Phys. Rev. Lett.}
{\bf 52}, 1072 (1982);\\
V. Kuzmin and M. Shaposnikov, {\it Phys. Lett.} {\bf B92 },115 (1980).

\item D. Chang and R.N. Mohapatra, {\it Phys. Rev.} {\bf D 32}, 1248
(1985).
\item D. G. Lee, R. N. Mohapatra, M. K. Parida and M. Rani, UMD-PP-94-117
(1994);\\
F. Acampora, G. Amelino-Camelia, F. Buccella, O. Pisanti, L. Rosa and
T. Tuzi, DSF Preprint-93/52 (1993).
\item Y. Totsuka, KEK Preprint (1992).
\item N. Ushida et. al., {\it Phys. Rev. Lett.} {\bf 57}, 2897 (1986).
\item C. Rubbia, CERN Preprint PPE-93-08 (1993).
\item L. Lavoura, {\it Phys. Rev.} {\bf D}  (1993).
\item R. N. Mohapatra and G. Senjanovic, {\it Zeit. fur Phys.},
{\bf C17}, 53 (1983);\\
R. Holman, G. Lazaridis and Q. Shafi, {\it Phys. Rev.} {\bf D27},995
(1983).
\item A. Joshipura, PRL preprint,PRL-TH/93/20 (1993);\\
P. Bamert and C. Burgess, McGill Preprint McGill-94/07 (1994);\\
S. T. Petcov and A. Smirnov, Trieste Preprint,(1993);\\
K. S. Babu and S. Pakvasa, {\it Phys. Lett.} {\bf B172}, 360 (1986).
\end{list}
\newpage

\end{document}